# A Theoretical Investigation of Relativistic Spintronics

Z. Y. Wang[1*], C. D. Xiong[1] and B. Chen[2]

[1]University of Electronic Science and Technology of China, Chengdu, Sichuan 610054, P.R. China

[2]School of Optics/CREOL, University of Central Florida, Orlando, FL 32816, USA

**Abstract**

Spintronics directly based on relativistic quantum mechanics is called as *relativistic spintronics*, which involves the study of active control and manipulation of 4D spin-tensor degrees of freedom via the electromagnetic field tensor. For future potential electronic devices with smaller size, the study of relativistic spintronics would be valuable. In this paper, we try to establish a general theoretical basis for relativistic spintronics, from which we have: 1) in relativistic spintronics, there have more plentiful contents for relativistic effects, e.g. the usual 3D spatial spin-orbit coupling is extended to *the 4D spin-orbit-tensor coupling*; 2) via the spin and *like-spin* degrees of freedom we can simultaneously make use of electric and magnetic field strengths to orientate an electron; 3) not only the spin quantum states of a moving electron but also those of a *motionless* electron can be affected by some particular *electrostatic* fields.

With the passage of time, the size of electronic devices becomes smaller and smaller, and the quantum-mechanics effects become more and more significant. Further, as the size of some future electronic devices becomes so small that the effects of relativistic quantum mechanics and quantum field theory, not only those of nonrelativistic quantum mechanics, play an important role in determining the properties of these electronic devices, and are crucial for their applications, we would have to investigate these device technologies based on relativistic quantum mechanics and quantum field theory rather than nonrelativistic quantum mechanics. In fact, though most microscopic interactions in condensed matter physics can be accurately described by nonrelativistic quantum mechanics, spin-orbit coupling that arises from the relativistic-quantum-mechanics effects described by the Dirac equation, is crucial for spintronics device applications [1], that is, one can make use of the spin-orbit coupling to manipulate electron spins by purely electric means [1-12].

In this paper, we try to establish a general theoretical basis for *relativistic spintronics* which is directly based on relativistic quantum mechanics. From which we can obtain more general perspective and conclusions with spin-orbit coupling only as one of illustrations. For simplicity we apply natural units of measurement ($\hbar = c = 1$) and let the metric tensor $g_{\mu\nu} = diag(1,-1,-1,-1)$, $\hat{p}_\mu = i\frac{\partial}{\partial x^\mu} \equiv i\partial_\mu$ denotes 4-momentum operators ($\mu, \nu = 0,1,2,3$).

In relativistic quantum mechanics, a free electron with mass $m$ is described by the Dirac equation for spin-$\frac{1}{2}$ particles:

$$(i\gamma^\mu \partial_\mu - m)\psi(x) = 0 \qquad (1)$$

where the four $4\times 4$ Dirac matrices $\gamma^\mu$ satisfy the algebra $\gamma^\mu\gamma^\nu + \gamma^\nu\gamma^\mu = 2g^{\mu\nu}$, the wave function $\psi(x)$ has four components and satisfies the transformation laws of a relativistic spinor. The Lagrange density that leads to Equ. (1) is Lorentz invariance, corresponding to which the conserved quantity is the total angular momentum tensor:

$$J^{\mu\nu} = x^\mu \hat{p}^\nu - x^\nu \hat{p}^\mu + S^{\mu\nu} \qquad (2)$$

where $S^{\mu\nu} = \frac{i}{4}[\gamma^\mu, \gamma^\nu] (\equiv \frac{i}{4}(\gamma^\mu\gamma^\nu - \gamma^\nu\gamma^\mu))$ is the 4-dimensional (4D) spin tensor. Let $\varepsilon^{ijk}$ denotes the totally anti-symmetric tensor with $\varepsilon^{123} = 1$ ($i,j,k=1,2,3$), one can show that $\vec{\Sigma} = (\Sigma_1, \Sigma_2, \Sigma_3)$ with $\Sigma_i = \frac{1}{2}\varepsilon_{ijk} S^{jk}$ is the usual spin matrices (as the generators for 3D spatial rotations). We call the generators $\vec{K} \equiv (S^{01}, S^{02}, S^{03})$ for Lorentz boosts as *spin-like matrices*. Conversely, the spin matrices $\vec{\Sigma}$ and *spin-like matrices* $\vec{K}$ can be assembled into a covariant anti-symmetric tensor, i.e. the 4D spin-tensor $S^{\mu\nu}$. Let $\vec{\sigma} = (\sigma_1, \sigma_2, \sigma_3)$ be the Pauli's matrix vector, where

$$\sigma_1 = \begin{pmatrix} 0 & 1 \\ 1 & 0 \end{pmatrix}, \quad \sigma_2 = \begin{pmatrix} 0 & -i \\ i & 0 \end{pmatrix}, \quad \sigma_3 = \begin{pmatrix} 1 & 0 \\ 0 & -1 \end{pmatrix} \qquad (3)$$

In terms of $\vec{\sigma}$, the spin matrices $\vec{\Sigma}$ and spin-like matrices $\vec{K}$ can be expressed as

$$\vec{\Sigma} = \frac{1}{2}\begin{pmatrix} \vec{\sigma} & 0 \\ 0 & \vec{\sigma} \end{pmatrix}, \quad \vec{K} = \frac{i}{2}\begin{pmatrix} 0 & \vec{\sigma} \\ \vec{\sigma} & 0 \end{pmatrix} \qquad (4)$$

Let $A^\mu = (\Phi, \vec{A})$ ($\mu = 0,1,2,3$) be a 4-potential of the external electromagnetic field ($\vec{A}$ being the vector potential and $A^0 = \Phi$ the scalar potential), $e$ the unit charge, and $F_{\mu\nu} \equiv \partial_\mu A_\nu - \partial_\nu A_\mu$ the electromagnetic field tensor. The first-order Dirac equations of an electron in $A^\mu$ can be transformed into second-order equations [13]:

$$[(i\partial^\mu - eA^\mu)(i\partial_\mu - eA_\mu) - m^2 - eS^{\mu\nu}F_{\mu\nu}]\psi(x) = 0 \qquad (5)$$

or ($\hat{p} = -i\nabla$),

$$[(i\frac{\partial}{\partial t} - e\Phi)^2 - (\hat{p} - e\vec{A})^2 - m^2 + 2e\vec{\Sigma}\cdot\vec{B} - 2e\vec{K}\cdot\vec{E}]\psi(x) = 0 \quad (6)$$

where the electric and magnetic field strengths $\vec{E} \equiv -\nabla\Phi - \frac{\partial}{\partial t}\vec{A}$, $\vec{B} \equiv \nabla\times\vec{A}$, they are assembled into a covariant anti-symmetric tensor, i.e. $F_{\mu\nu}$. The $i = \sqrt{-1}$ factor in $\vec{K}$ (expressed by Equ. (4)) is necessary to assure the Hermiticity of $\vec{K}\cdot\vec{E}$ in Equ. (6).

Conventionally electronics only sensitive to electron's charge, spin degree of freedom ignored. Spintronics, a new research field develop in recent years, is based on the up ($\uparrow$) or down ($\downarrow$) spin of carriers rather than on electrons or holes as in traditional semiconductor electronics, and involves the study of active control and manipulation of spin degrees of freedom in solid-state systems [14]. Further, as one extends spintronics mainly based on nonrelativistic quantum mechanics to *relativistic spintronics* directly based on relativistic quantum mechanics, not only spin degrees of freedom related to the spin matrices $\vec{\Sigma}$, but also *spin-like* degrees of freedom related to the spin matrices $\vec{K}$ would enter into our consideration. A complete and accurate (i.e. without any approximation) description for the related interactions in *relativistic spintronics* is represented by $eS^{\mu\nu}F_{\mu\nu}$ in Equ. (5). Therefore, *relativistic spintronics* involves the study of active control and manipulation of 4D spin-tensor $S^{\mu\nu}$ degrees of freedom via the electromagnetic field tensor $F_{\mu\nu}$. In view of which, we will study the interactions $-eS^{\mu\nu}F_{\mu\nu} \equiv \hat{H}_{SF}$ in Equ. (5) in a general way. Consider that the Dirac equation in the external fields $A^\mu$:

$[\gamma^\mu(\hat{p}_\mu - eA_\mu) - m]\psi(x) = 0$, that is $\psi = \frac{\gamma^\mu(\hat{p}_\mu - eA_\mu)}{m}\psi$, we have (see Equ. (5),

$\alpha, \beta, \lambda = 0, 1, 2, 3$ )

$$\hat{H}_{SF}\psi = -eS^{\alpha\beta}F_{\alpha\beta}\psi = -\frac{e}{m}S^{\alpha\beta}F_{\alpha\beta}[\gamma^{\lambda}(\hat{p}_{\lambda} - eA_{\lambda})]\psi \quad (7)$$

Let $\tau = \sqrt{x^{\mu}x_{\mu}}$ denote the Lorentz-invariant proper time, then $\partial_{\mu} = \frac{\partial}{\partial x^{\mu}} = \frac{x_{\mu}}{\tau}\frac{\partial}{\partial \tau}$.

Using $a^{\mu}\gamma_{\mu}b^{\nu}\gamma_{\nu} = a^{\mu}b_{\mu} - 2ia^{\mu}b^{\nu}S_{\mu\nu}$, $S^{\mu\nu} = \frac{i}{4}[\gamma^{\mu}, \gamma^{\nu}]$ and the Lorentz gauge condition $\partial_{\mu}A^{\mu} = 0$, we obtain

$$\hat{H}_{SF}\psi = -eS^{\alpha\beta}F_{\alpha\beta}\psi = [\frac{ie}{m}\gamma_{\mu}(\partial_{\nu}A^{\mu})(\hat{p}^{\nu} - eA^{\nu}) + \frac{e}{m}\frac{1}{\tau}\frac{\gamma^{\mu}\partial A_{\mu}}{\partial \tau}S_{\alpha\beta}\hat{L}_{A}^{\alpha\beta}]\psi \quad (8)$$

where $\hat{L}_{A}^{\mu\nu} \equiv x^{\mu}(\hat{p}^{\nu} - eA^{\nu}) - x^{\nu}(\hat{p}^{\mu} - eA^{\mu})$ is the 4D orbit-angular-momentum tensor in the external fields $A^{\mu}$. We refer to $\frac{e}{m}\frac{1}{\tau}\frac{\gamma^{\mu}\partial A_{\mu}}{\partial \tau}S_{\alpha\beta}\hat{L}_{A}^{\alpha\beta}$ in Equ. (8) as *the 4D spin-orbit-tensor coupling*, via which the degrees of freedom of the 4D spin and orbit angular momentum tensors are coupled, where the usual 3D spatial spin-orbit coupling is contained. As for *relativistic spintronics*, Equ. (8) describes all the interactions related to 4D spin-tensor $S^{\mu\nu}$ degrees of freedom, without any approximation.

Therefore, the most general starting point of the theoretical analysis for *relativistic spintronics* has two equivalent forms: Equ. (8) and the last term on the right-hand side of Equ. (5) (or the last two terms on the right-hand side of Equ. (6)).

We now come back to the last two terms on the right-hand side of Equ. (6). When the electric and magnetic field strengths $\vec{E}$ and $\vec{B}$ as two external fields satisfy $\vec{E} = k\vec{B}$ ($k$ is a constant), we have $[\vec{\Sigma}\cdot\vec{B}, \vec{K}\cdot\vec{E}] = 0$, that is, $2e\vec{\Sigma}\cdot\vec{B}$ and $-2e\vec{K}\cdot\vec{E}$ have common eigenstates. This means that we can simultaneously make use of the electric and magnetic field strengths $\vec{E}$ and $\vec{B}$ to orientate an electron (e.g. the directions of $\vec{E}$ and $\vec{B}$ are perpendicular to a two-dimensional plane where the polarization electrons are localized and orientated). In other words, simultaneously make use of the spin and *like-spin* degrees of freedom to orientate an electron (we will give a more detailed discuss later). This offers a way to strengthen the control and manipulation for the spin degrees of freedom of electron. Various disposal and approximation methods of the term $eS^{\mu\nu}F_{\mu\nu}$ in Equ. (5) open attractive and multiplicate possibilities for control of the 4D-spin-tensor degrees of freedom of electron, where including purely electrical manipulation of moving or *resting* electron spins (we will discuss later). So there is great promise for the application of relativistic spintronics in future potential electronic devices with smaller size.

In order to understand what is the physical meanings of making use of the spin and *like-spin* degrees of freedom to orientate an electron, as an example, let $\psi = \psi' \exp(-im)$, $\psi' = \begin{pmatrix} \varphi \\ \chi \end{pmatrix}$, i.e., the four-component spinor $\psi'$ is decomposed into two two-component spinors $\varphi$ and $\chi$. In the nonrelativistic and weak external fields limit, for the four-component spinor $\psi'$ belonging to positive (negative) energy states, the upper (lower) two components $\varphi$ ($\chi$) become large compared to the lower (upper) two components $\chi$ ($\varphi$). In the approximation $\left|i\frac{\partial}{\partial t}\psi'\right|$, $|e\Phi\psi'| \ll m|\psi'|$ and $\left|i\frac{\partial}{\partial t}\Phi\right| \ll m|\Phi|$, Equ. (6) becomes

$$i\frac{\partial}{\partial t}\psi' = [\frac{(\hat{p}-e\vec{A})^2}{2m} + e\Phi - g_s\frac{e}{2m}\vec{\Sigma}\cdot\vec{B} + g_s\frac{e}{2m}\vec{K}\cdot\vec{E}]\psi' \qquad (9)$$

where the g-factor $g_s = 2$. The electron will therefore behave as though it has a magnetic moment operator $g_s \frac{e}{2m}\vec{\Sigma}$ and an electric moment operator $g_s \frac{e}{2m}\vec{K}$. As well known, an observable is represented by the expectation value of a dynamical operator or the squared norm of a state vector, instead of by the dynamical operator or state vector itself. Especially, in the Schrödinger picture, the wave functions (or state vectors) serve as the carrier of the relativistic effects of the purely numerical matrix operators $\vec{\Sigma}$ and $\vec{K}$. Judged by the relation between the averages of $g_s \frac{e}{2m}\vec{\Sigma}$ and $g_s \frac{e}{2m}\vec{K}$ in the state vector $\psi'$, one can show that the magnetic moment, related to the spin matrices $\vec{\Sigma}$ (the purely spatial components of the 4D spin-tensor $S^{\mu\nu}$), is nonzero in the rest frame of electron (say, the magnetic moment is *intrinsic*); whereas the electric moment, related to *spin-like* matrices $\vec{K}$ (the space-time components of the 4D spin-tensor $S^{\mu\nu}$), is the relativistic effect of the magnetic moment (say, the electric moment is *induced*). For example, using $\psi' = \begin{pmatrix} \varphi \\ \chi \end{pmatrix}$ and Equ. (4), it is easily to show that, in the nonrelativistic and weak external fields limit, $(\vec{\Sigma}\cdot\vec{B} - \vec{K}\cdot\vec{E})\psi' \to (\vec{\sigma}\cdot\vec{B} - \frac{v}{c}\vec{\sigma}\cdot\vec{E})\varphi$, where $v$ and $c(=1)$ are the velocities of the electron and light in the vacuum, respectively.

Let us emphasize that the so-called *intrinsic* or *permanent* dipole electric moment of electron [15] that people are searching nowadays (in order to provide an interpretation for the violation of time reversal invariance) is conceptually distinct from the *induced* electric moment of electron mentioned here. The *intrinsic* electric dipole moment implies a charge distribution with respect to the center of a particle, while the *induced* electric moment originates from the relativistic-quantum-mechanics effects that have not a counterpart in classical mechanics, and the interaction between the *induced* electric moment and electric fields does not break time reversal symmetry.

Furthermore, by applying $\psi' = \begin{pmatrix} \varphi \\ \chi \end{pmatrix}$, Equ. (9) becomes

$$i\frac{\partial}{\partial t}\varphi = [\frac{(\hat{p}-e\vec{A})^2}{2m} + e\Phi - \frac{e}{2m}\vec{\sigma}\cdot\vec{B}]\varphi + i\frac{e}{2m}\vec{\sigma}\cdot\vec{E}\chi$$
$$i\frac{\partial}{\partial t}\chi = [\frac{(\hat{p}-e\vec{A})^2}{2m} + e\Phi - \frac{e}{2m}\vec{\sigma}\cdot\vec{B}]\chi + i\frac{e}{2m}\vec{\sigma}\cdot\vec{E}\varphi \quad (10)$$

One can easily show the Hermiticity of $i\frac{e}{2m}\vec{\sigma}\cdot\vec{E}$ in Equ. (10) because both the $\vec{E}$ and $i\vec{\sigma}$ *operators* are anti-Hermitian and $[\vec{E}, i\vec{\sigma}] = 0$. Then the interaction $i\frac{e}{2m}\vec{\sigma}\cdot\vec{E}$, related to *spin-like* degrees of freedom, results in the coupling of two-component spinors $\varphi$ and $\chi$. As we known, $\varphi$ is the *large* components while $\chi$ the *small* components for electron but conversely for positron, that is, in the interchange $\varphi \leftrightarrow \chi$, we have electron $\leftrightarrow$ positron. As for a wave packet of electron, the relative intensities of its negative-energy and positive-energy components are proportional to $|\chi|^2/|\varphi|^2$. Judged by Equ. (4), (9) and (10), in contrast with spin degrees of freedom that are related to the spin matrices $\vec{\Sigma}$ and described by spin-$\uparrow$ and spin-$\downarrow$ states, *spin-like* degrees of freedom related to the spin matrices $\vec{K}$ involve positive-energy and negative-energy states, or, involve the particle-antiparticle degrees of freedom, which accords with the facts that the wave function $\psi(x)$ of electron has four components rather than two components, and a superposition of plane waves of positive as well as of negative energy is necessary to obtain a wave packet of electron [16].

By applying Equ. (10) and the approximate relation of $\varphi$ and $\chi$, one can easily show that the spin-orbit coupling and the Darwin interactions are contained in $i\frac{e}{2m}\vec{\sigma}\cdot\vec{E}$. More general, we consider an approximation of the exact solution of Equ. (5) or (6) with the help of the Foldy-Wouthuysen transformation [16], let $\Phi=\Phi(r)$, $r=\sqrt{\bar{x}^2+\bar{y}^2+\bar{z}^2}$, $\hat{L}=\vec{r}\times\hat{p}$ the 3D spatial angular momentum and $\vec{E}=-\nabla\Phi(r)=-\frac{\vec{r}}{r}\frac{d}{dr}\Phi(r)$, for the moment the interactions $eS^{\mu\nu}F_{\mu\nu}$ become, approximatively,

$$\hat{H}'_{SF}=-\frac{e}{m}\gamma^0(\vec{\Sigma}\cdot\vec{B})-\frac{ie}{4m^2}\vec{\Sigma}\cdot(\nabla\times\vec{E})-\frac{e}{8m^2}\nabla\cdot\vec{E}-\frac{e}{2m^2}\frac{1}{r}\frac{\partial\Phi}{\partial r}(\vec{\Sigma}\cdot\hat{L}) \quad (11)$$

where the third term on the right-hand side of Equ. (11) is the so-called Darwin term and last term represents the spin-orbit coupling, Equ. (11) is one of the specific approximations of Equ. (8).

Let $\psi_+=\varphi+\chi$ and $\psi_-=\varphi-\chi$, Equ. (10) becomes

$$i\frac{\partial}{\partial t}\begin{pmatrix}\psi_+\\\psi_-\end{pmatrix}=\hat{H}\begin{pmatrix}\psi_+\\\psi_-\end{pmatrix} \quad (12)$$

Where

$$\hat{H}=\begin{pmatrix}\hat{H}_1+i\frac{e}{2m}\vec{\sigma}\cdot\vec{E} & 0\\ 0 & \hat{H}_1-i\frac{e}{2m}\vec{\sigma}\cdot\vec{E}\end{pmatrix}, \quad \hat{H}_1=\frac{(\hat{p}-e\vec{A})^2}{2m}+e\Phi-\frac{e}{2m}\vec{\sigma}\cdot\vec{B} \quad (13)$$

For simplicity, we consider a special case: let $\vec{B}=\nabla\times\vec{A}=0$, $\vec{E}=-\nabla\Phi-\frac{\partial}{\partial t}\vec{A}$ with $\frac{\partial\vec{E}}{\partial t}=0$, and $\hat{H}_1=\frac{(e\vec{A})^2}{2m}+e\Phi$ (namely, a *rest* electron in a purely *electrostatic* field), then $[\hat{H}_1,i\frac{e}{2m}\vec{\sigma}\cdot\vec{E}]=0$, this means that $\hat{H}_1$ and $i\frac{e}{2m}\vec{\sigma}\cdot\vec{E}$ can be diagonalized together and have common eigenstates, let

$$\hat{H}_1\begin{pmatrix}\psi_+\\\psi_-\end{pmatrix}=\varepsilon_1\begin{pmatrix}\psi_+\\\psi_-\end{pmatrix}, \quad i\frac{e}{2m}\vec{\sigma}\cdot\vec{E}\begin{pmatrix}\psi_+\\\psi_-\end{pmatrix}=\varepsilon'\begin{pmatrix}\psi_+\\\psi_-\end{pmatrix} \quad (14)$$

where we let $\varepsilon' > 0$ without loss of generality, then

$$\hat{H}\begin{pmatrix}\psi_+\\\psi_-\end{pmatrix}=\begin{pmatrix}\varepsilon_1-\varepsilon' & 0\\ 0 & \varepsilon_1+\varepsilon'\end{pmatrix}\begin{pmatrix}\psi_+\\\psi_-\end{pmatrix} \qquad (15)$$

that is, as $\vec{E}=0$, $\psi_+ = \varphi+\chi$ and $\psi_- = \varphi-\chi$ are two degeneration states of $\hat{H}$ with the same eigenvalue $\varepsilon_1$; as $\vec{E} \neq 0$, the degeneracy of the two-fold multiplet is broken by the electrostatic field $\vec{E}$, which is analogous to the Zeeman effect, but here corresponds to the interaction $i\frac{e}{2m}\vec{\sigma}\cdot\vec{E}$ (associated with *spin-like* degrees of freedom) instead of the interaction $\frac{e}{2m}\vec{\sigma}\cdot\vec{B}$ (associated with spin degrees of freedom). Therefore, a purely electrostatic field can affect the *spin-like* quantum states of a rest electron by means of the interaction $i\frac{e}{2m}\vec{\sigma}\cdot\vec{E}$, while the Refs. [4-5] show that a purely electrostatic field can affects the spin quantum states of a moving electron, i.e. spin and motional degrees of freedom are coupled via spin-orbit coupling, which due to the fact that an electric field $\vec{E}$ in lab frame causes a magnetic field $\vec{B}=\frac{1}{c^2}\vec{E}\times\vec{v}$ in rest frame of moving electron with the velocity $\vec{v}$. In view of the fact that the *induced* electric moment is the relativistic effect of the *intrinsic* magnetic moment, whether the interaction of the purely electrostatic field and rest electron mentioned here or spin-orbit coupling is attributed to the relativistic-quantum effects. In a word, Equ. (15) shows that some specific *electrostatic* fields can also affect the spin quantum states of a *rest* electron by means of the interaction $i\frac{e}{2m}\vec{\sigma}\cdot\vec{E}$ that related to *spin-like* degrees of freedom.

On the one hand, in the relativistic or strong-external-field limit, two components $\chi$ of the bispinor $\psi'=\begin{pmatrix}\varphi\\\chi\end{pmatrix}$ of electron can not be ignored, and the interactions related to *spin-like* degrees of freedom ought to be taken into account. On the other hand, owing to the Heisenberg uncertainty principle, if the width of the wave packet of an electron is compressed to a size of about a Compton wavelength of the electron, the partial waves of negative energies have an appreciable effect and the interactions related to *spin-like* degrees of freedom must be considered. All these cases may hold true for future potential electronic devices with smaller size, and hence the study of the concrete applications of *relativistic spintronics* would be valuable.